\documentclass[10pt,preprint2]{aastex}
\shorttitle{Near-IR Spectral Library}
\shortauthors{Ivanov et al.}

\begin{document}

\title{A Medium-Resolution Near-Infrared Spectral Library\\ 
of Late Type Stars: I.}
\author{Valentin D. Ivanov}
\affil{European Southern Observatory, Karl-Schwarzschild-Str. 2,
   D-85748 Garching bei M\"unchen, Germany}
\email{vivanov@eso.org}
\author{Marcia J. Rieke, Charles W. Engelbracht, Almudena
   Alonso-Herrero\footnote{Present address: Departamento de 
Astrof\'{\i}sica Molecular e Infrarroja, IEM, Consejo Superior 
de Investigaciones Cient\'{\i}ficas, Serrano 113b, 28006 Madrid, 
Spain}, George H. Rieke}
\affil{Steward Observatory, The University of Arizona, Tucson, AZ85721,
   U.S.A.}
\email{mrieke, chad, grieke, aalonso@as.arizona.edu}
\author{Kevin L. Luhman}
\affil{Harvard-Smithsonian Center for Astrophysics, 60 Garden Street,
   Cambridge, MA 02138, U.S.A.}
\email{kluhman@cfa.harvard.edu}

\begin{abstract}
We present an empirical infrared spectral library of medium 
resolution (R$\approx$2000-3000) $H$ ($1.6\,\mu$m) and $K$ 
($2.2\,\mu$m) band spectra of 218 red stars, spanning a range 
of [Fe/H] from $\sim$$-$2.2 to $\sim$+0.3. The sample includes 
Galactic disk stars, bulge stars from Baade's window, and 
red giants from Galactic globular clusters. We report the 
values of 19 indices covering 12 spectral features measured 
from the spectra in the library. Finally, we derive 
calibrations to estimate the effective temperature, and 
diagnostic relationships to determine the luminosity classes of 
individual stars from near-infrared spectra.

This paper is part of a larger effort aimed at building a 
near-IR spectral library to be incorporated in population 
synthesis models, as well as, at testing synthetic stellar 
spectra.
\end{abstract}


\section{Introduction}

Over the last two decades, evolutionary synthesis modeling has 
become a common tool to study unresolved stellar populations of 
galaxies in the optical and UV passbands 
\citep{bru93,wor94a,vaz99,lei99}. The heavy obscuration in 
starburst galaxies -- often with A$_V$$\geq$5 mag~ 
\citep{eng97} -- requires an expansion of these models into the near
infrared (IR) domain, because the extinction there is reduced 
tenfold in comparison with the optical region: A$_K$/A$_V$=0.112
\citep{rie85}. Therefore, it is not surprising that many studies of 
embedded stellar populations in galaxies have been conducted in the 
near-IR \citep{rie80,rie93}. 

IR evolutionary synthesis models based on synthetic spectra
\citep{ori93,kur94} generally have difficulties reproducing
broad-band colors because of the complicated opacity calculations 
in the near-IR. Empirical libraries based predominantly on bright and 
nearby Milky Way stars 
\citep{kle86,lan92,wal96,wal97,mey98,wal00} are limited to near 
solar metallicity. At the same time, realistic galaxy modeling 
requires metallicities ranging from [Fe/H]=$-$1.76 for the most 
metal-poor galaxy known I\,Zw\,18 \citep{alo03}, to supersolar 
values, measured for at least for some giant elliptical galaxies 
\citep{cas96}.

The properties of the most prominent near-IR spectral libraries
available in the literature are summarized in Table\,\ref{TblLibRev}.
For a review of the earlier work see \citet{mer79}. This summary 
shows that despite the sizable quantity of available stellar 
spectra, up until now there was no uniform near-IR dataset with 
high signal-to-noise and resolution, covering the entire range of 
spectral classes, luminosity, and metallicity necessary to carry out
evolutionary population synthesis in the near-IR. The deficit of 
some types of stars such as metal-poor super giants is understandable
in the context of the Milky Way star formation history, but the lack 
of many other types is rectifiable. We concentrate on metal features 
that have equivalent widths in a typical starburst galaxy larger 
than 1\AA\, -- as they can be measured reliably in  spectra with 
signal-to-noise ratios of $\sim$30-50 \citep{eng97}.

A second application of our library, albeit no less important, is 
the analysis of individual stars hidden behind A$_V$$\geq$10 mag of 
visual extinction. A typical case for such a population is 
represented by the Arches cluster member starss in the Galactic 
Center region \citep{nag93}.

Here we describe the observations and the sample of an empirical 
near-IR stellar library that is designed to meet 
the requirements of population synthesis. The main 
advantage in comparison with previous work is the expanded 
metallicity coverage. We also present a few diagnostics methods to 
derive parameters of individual stars. The evolutionary population 
model will be published in a subsequent paper.
The next section describes the observations and the data reduction 
technique. Sec.\,3 summarizes the sample selection. Spectral indices 
are defined in Sec.\,4, and some diagnostic applications for 
parameters of individual stars are considered in Sec.\,5. In 
Sec.\,6. we give the summary.

\section{Observations and Data reduction}

The near-IR spectra were taken from 1995 to 1999 mainly at the
Steward Observatory 2.3m Bok telescope on Kitt Peak. Some stars
were observed at the original 4.5m MMT and at the Steward
Observatory 1.55m Kuiper telescope. We used FSpec \citep{wil93}, 
a cryogenic long slit near-IR spectrometer utilizing a NICMOS3 
256x256 array \citep{koz93}. The majority of the stars were 
observed with a 600 lines mm$^{-1}$ grating, corresponding to 
a spectral resolution $R$$\approx$2000-3000.
This is the highest useful spectral resolution for studies of 
stellar populations in external galaxies where the intrinsic 
velocity dispersion smooths out the integrated spectra.
The rest were observed with a 300 lines mm$^{-1}$ grating and
$R$$\approx$1000-1500. The slit widths were 2.4 arcsec at the 2.3m,
1.2 arcsec at the MMT, and 3.6 arcsec at the 1.55m. The plate scales
were 1.2, 0.6, and 1.8 arcsec pixel$^{-1}$, respectively. The
limited physical size of the near-IR array required the acquisition
of spectra at 10-12 grating positions, corresponding to different
central wavelengths ($\lambda_c$) to cover the entire $H$ and $K$ 
atmospheric windows with sufficient overlap. Usually, one of the 
following two combinations of $\lambda_c$ was used: 
1.54, 1.62, 1.70, 
2.06, 2.13, 2.19, 2.245, 2,31, 2.37, 2.42 $\,\rm\mu m$, 
or 
1.50, 1.57, 1.64, 1.71, 
2.05, 2.10, 2.15, 2.20, 2.25, 2.30, 2.35, 2.40 $\,\rm\mu m.$
The log of observations is given in Table~\ref{TblObsLog}.

The observing strategy for a single setting included nodding the
telescope to obtain spectra at 4 (at the MMT) or 6 (at the 2.3m and
1.55m telescopes) different positions along the slit, and integrating 
at each position for $3-20$ seconds, depending on the apparent 
brightness of the object and the sky background. This was necessary 
in order to: 
(i) carry out the sky emission subtraction and account for the sky 
background variations by having a sky taken just before and after 
the science exposures;
(ii) improve the pixel sampling, flat fielding, and bad pixel 
correction by having the object placed on multiple positions on 
the array. 

Next, we repeated the same procedure for a standard star at similar 
airmass (airmass difference \,$\leq$\,0.1-0.15) to correct 
for the atmospheric absorption. Then we changed the grating setting, 
obtained a sequence of $4-6$ spectra of the standard, moved the 
telescope back to the object, and repeated the same procedure at the 
new grating setting. Occasionally, if two or three target stars were 
available nearby on the sky, we increased 
the observing efficiency by using the same telluric standard for all 
of them.

The spectra were reduced using $IRAF$\footnote{IRAF is distributed
by the National Optical Astronomy Observatories, which are operated
by the Association of Universities for Research in Astronomy, Inc.,
under cooperative agreement with the National Science Foundation.}
tasks written specifically for FSpec. An average of the sky
backgrounds taken immediately before and after each object was
subtracted to remove simultaneously the sky emission lines, and the
dark current and bias level. Dark-subtracted illuminated dome flats
taken at the same central wavelengths as the science exposures were
applied to correct for pixel-to-pixel variations. The known bad
pixels were masked out. Object images were shifted (using centroid
fitting across the continuum), and median combined to produce a 
single two-dimensional spectrum. The large number of object images 
allowed us to reject any remaining bad pixels and cosmic rays.
One-dimensional spectra were extracted by fitting a polynomial of
order 3-5 to the continuum in the two-dimensional image with the
$IRAF$ task $apall$. The spatial width of the extraction apertures
was usually 3-5 pixels, depending on the scale and the seeing.

In most cases the object spectra were divided by spectra of solar
analog stars observed at the same airmass, and then multiplied by 
the solar spectrum \citep{liv91} to remove the effects of the 
photospheric absorption \citep{mai96}. The standard stars were not 
always exactly G2 dwarfs (usually ranging from F8\,V to G3\,V), and 
the true shape of the continuum had to be restored by multiplying 
the spectra by the ratio of black bodies with the standard star and 
solar effective temperatures. In the cases we chose nearby A stars
for telluric standards as an alternative of the solar analogs, we 
multiplied by another A star spectrum, already corrected for the 
photospheric absorption with a G2\,V star. 
If an A type standard was used to correct only a $K$-band spectrum, 
we removed the $\rm Br\gamma$ feature by interactively fitting and 
subtracting a Gaussian with the task {\it splot}.
We carried out the wavelength calibration using  OH airglow lines
\citep{oli92} complemented if necessary with Ne-Kr lamp spectra.

The spectrum at each individual setting was divided by second or 
third order polynomial continuum fits and combined with the rest 
of the settings to construct a single $HK$ spectrum. Finally, we 
multiplied the spectrum by a black body with stellar effective 
temperature corresponding to the spectral type \citep{sch82,str92}.
Unfortunately, the normalization procedure removes the broad vapor
absorption features in the coolest stars but this loss is unavoidable 
because
the spectral extent of a single setting is only $0.075-0.095\,\mu$m
and does not provide sufficient coverage for a reliable estimate of the
continuum shape. Furthermore, in many cases the true continuum shape
was distorted by the imperfect spectral type match of the telluric
standards, as discussed above.
The deviation between the constructed continuum shape and the real 
one is most severe for late M stars, where the molecular absorption is 
the strongest. A possible way to alleviate this problem completely is 
to impose on our spectra empirical continuum shapes taken from lower 
resolution spectra \citep{lan92,lan00}. We intend to explore this 
avenue in the future.

\begin{figure}
\epsscale{0.95}
\plotone{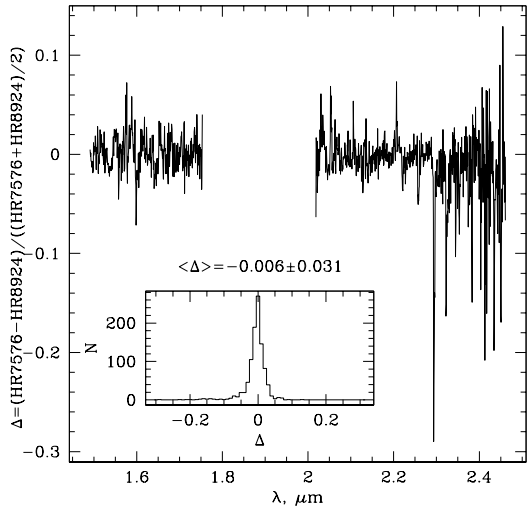}
\caption{Data quality: difference $\Delta$ between the spectra 
of two K3\,III stars -- HR\,7576 and HR\,8924 -- 
with similar abundances, normalized by the 
average of the two spectra as a function of wavelength (in 
arbitrary flux units). The inset 
shows the histogram of the differences. The means and the 
standard deviation of the distribution are also shown.
\label{Fig1a} }
\end{figure}

The error analysis of spectra that consist of ``pieces'' taken at
different times, and often on different nights, is particularly
complicated because of the large variations of the signal-to-noise
ratio (S/N) across the individual spectra.
The uncertainties are dominated by systematic errors due to: 
(i) sky emission and transparency variations with wavelength, 
and
(ii) temporal changes in the observing conditions.
An example is given in Fig.\,\ref{Fig1a}, where the difference
between the ``final'' spectra of two different K3\,III stars with
near solar metallicity is shown. The difference $\Delta$
has been divided by the average of the two spectra. The inset 
shows the distribution of these differences. Their average 
difference $<$$\Delta$$>$ is statistically 
indistinguishable from zero, and the value of the standard deviation 
$\sigma$ suggests that each spectrum has an average S/N$\sim$50. 
At the same time, the photon statistics suggests S/N$\sim$150. 
The inconsistency is largely due to the variation of the sky 
background and transmission. 
Therefore, a single signal-to-noise ratio does not represent well the 
quality of the data. Furthermore, the measurements of individual 
features of interest include additional systematic uncertainties, 
such as the continuum placement during the continuum normalization, 
and differences between the stars used for the telluric correction 
and the Sun.

\begin{figure}
\epsscale{0.95}
\plotone{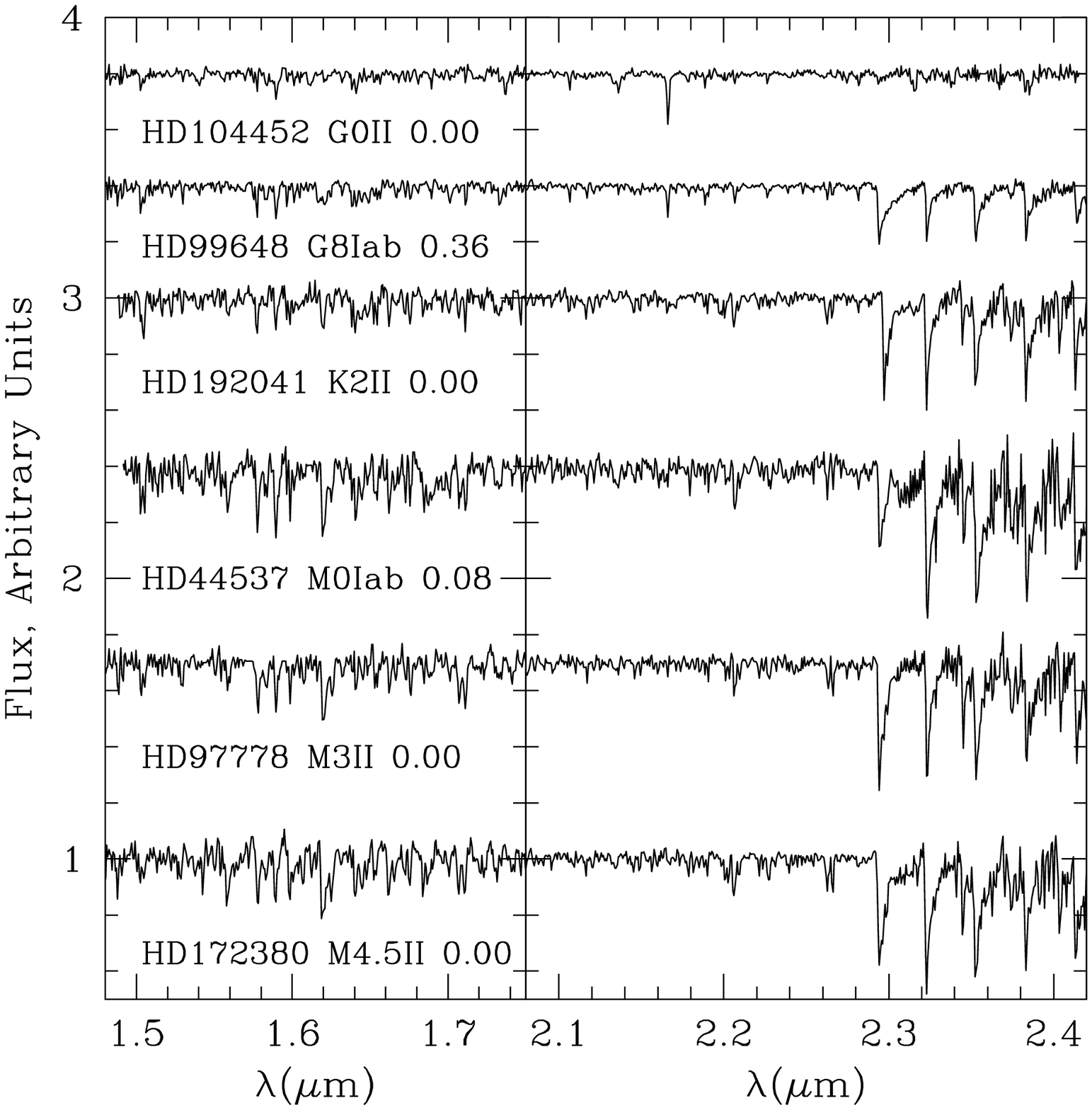}
\caption{A subset of H and K spectra of supergiants. 
The names of the stars, spectral types, and [Fe/H]
are indicated. The spectra are continuum divided and shifted
vertically for display purposes by adding (from bottom to top):
0.0, 0.7, 1.4, 2.0, 2.4, and 2.8.
\label{Fig4a} }
\end{figure}

\begin{figure}
\epsscale{0.95}
\plotone{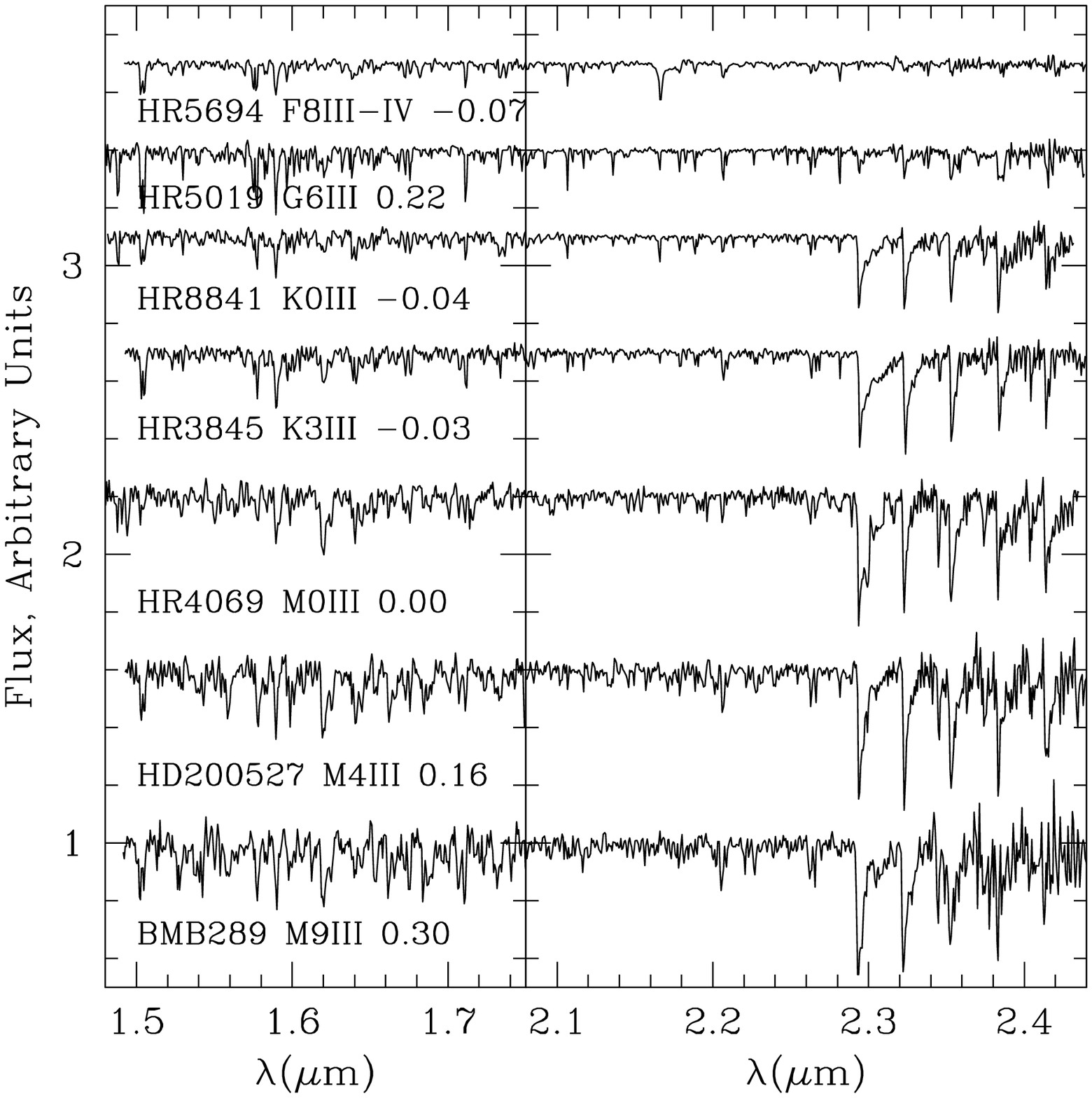}
\caption{A subset of H and K spectra of giants. 
The names of the stars, spectral types, and [Fe/H]
are indicated. The spectra are continuum divided and shifted
vertically for display purposes by adding (from bottom to top):
0.0, 0.6, 1.2, 1.7, 2.1, 2.4, and 2.7.
\label{Fig4b} }
\end{figure}

\begin{figure}
\epsscale{0.95}
\plotone{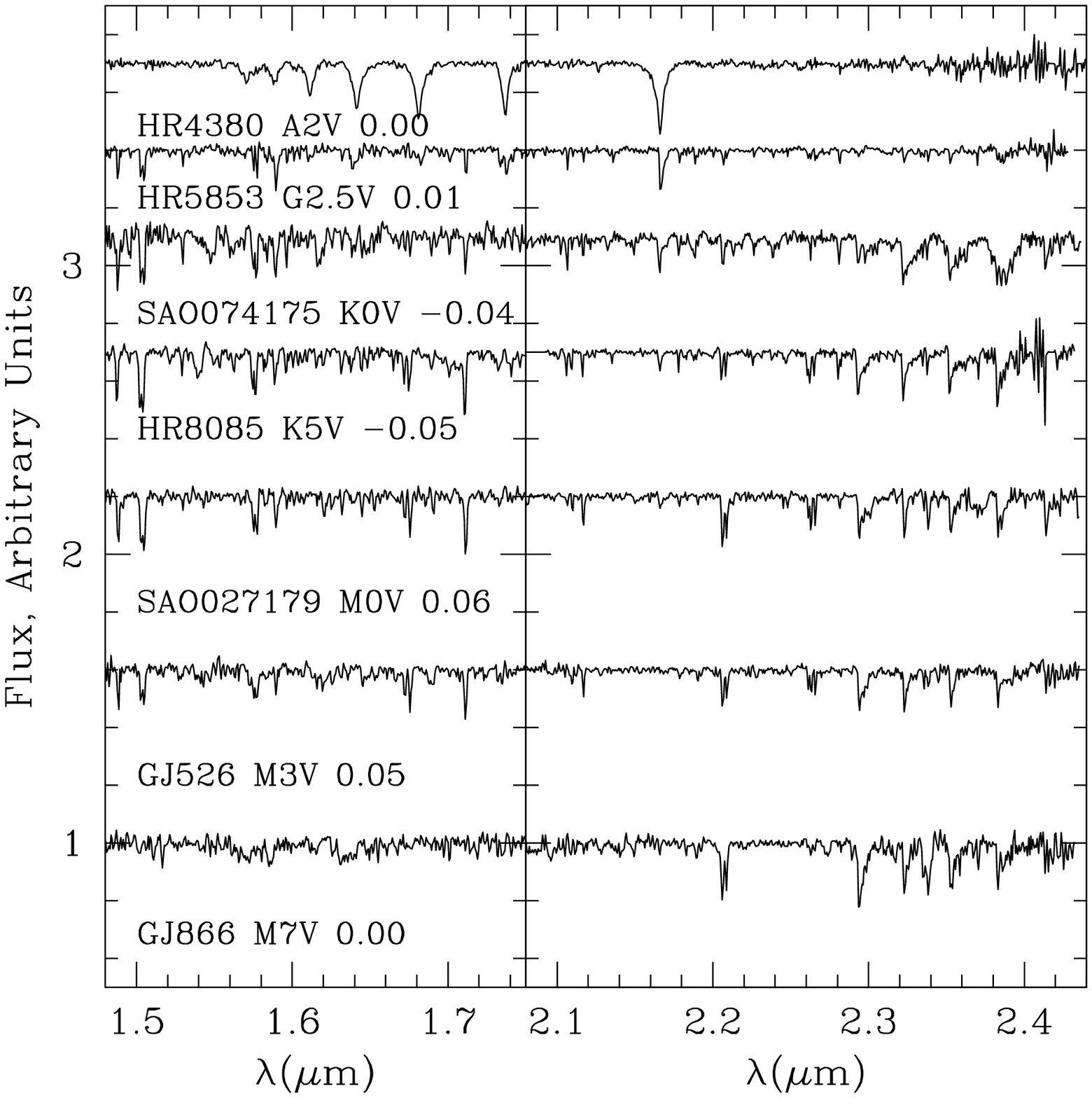}
\caption{A subset of H and K spectra of main sequence stars. 
The names of the stars, spectral types, and [Fe/H]
are indicated. The spectra are continuum divided and shifted
vertically for display purposes by adding (from bottom to top):
0.0, 0.6, 1.2, 1.7, 2.1, 2.4, and 2.7.
\label{Fig4c} }
\end{figure}

\begin{figure}
\epsscale{0.95}
\plotone{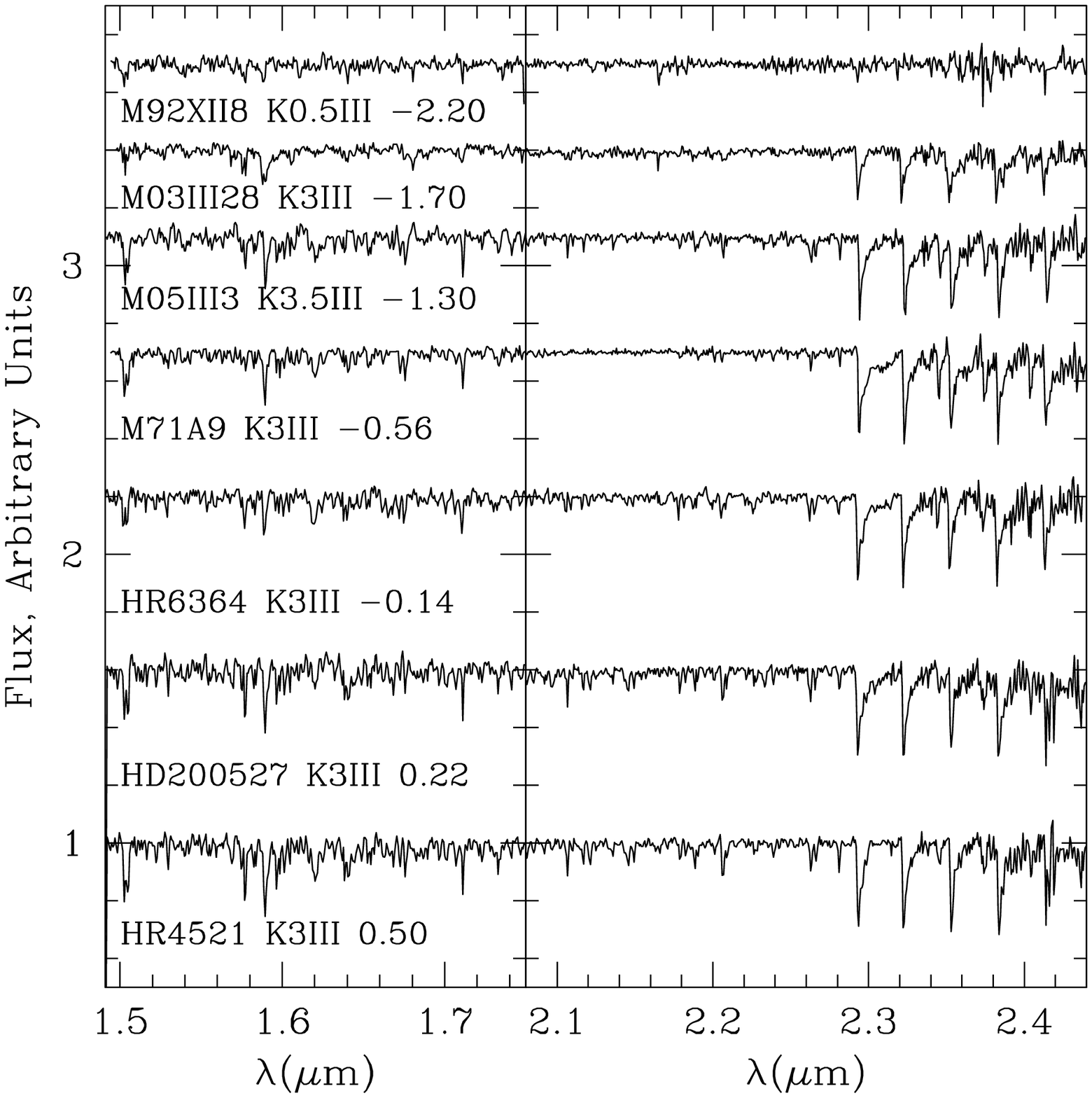}
\caption{A subset of H and K spectra for a selection of K giants 
spanning a range of metallicities. 
The names of the stars, spectral types, and [Fe/H]
are indicated. The spectra are continuum divided and shifted
vertically for display purposes by adding (from bottom to top):
0.0, 0.6, 1.2, 1.7, 2.1, 2.4, and 2.7.
\label{Fig4d} }
\end{figure}

The best technique to estimate the actual signal-to-noise ratio is to
measure the noise locally over a ``clean'' continuum region near
the spectral feature of interest. We recommend this method for
applications concerning parameters of individual stars. Our
study shows that for statistical work the uncertainties of indices
can be approximated as $\sigma(Index)=0.2\times Index,$ with a
lower limit of $\sigma(Index)=0.02$ mag. Representative subsets of
our spectra are shown in Fig.\,\ref{Fig4a}, \ref{Fig4b}, 
\ref{Fig4c}, and \ref{Fig4d}.

\section{Sample Selection and Stellar Parameters}

We have assembled an IR spectral library of 218 stars. The majority 
of them have photospheric parameters available from the literature. 
The main criterion to include various types of stars was to 
populate the {\it effective temperature -- surface gravity -- 
abundance} space necessary to model the stellar populations of 
starburst galaxies. Special attention was paid to observing red 
supergiants, which dominate the near-IR flux of these galaxies. 

\begin{figure}
\epsscale{0.95}
\plotone{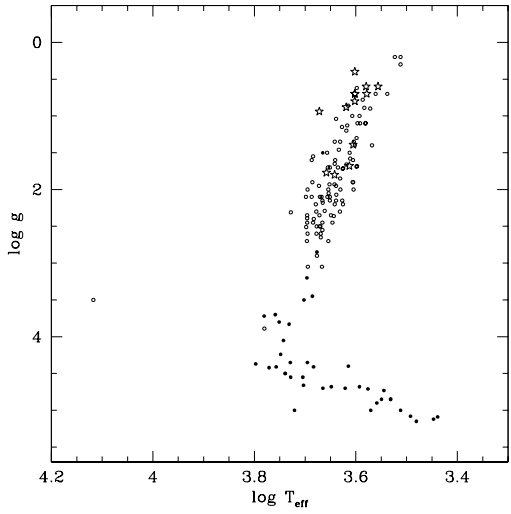}
\caption{Content of the library: distribution of the library stars
on the surface gravity log\,$g$ versus effective temperature 
$\rm T_{eff}$ plane. Star symbols are supergiants, circles are giants, and 
solid dots are sub-giants and dwarfs. The ``anomalous'' sub-giant 
at log\,$g$=1.5 is extremely metal poor, with [Fe/H]=$-$2.67.
\label{Fig1} }
\end{figure}

The largest fraction of stars in our library was selected from 
the Lick group sample \citep{wor94b}, assuring that they have high 
quality optical spectra and known spectral type, surface gravity,  
and metallicity. However, the Lick library was aimed at modeling 
galaxies with old stellar populations -- ellipticals and bulges of
spirals, -- and therefore, is mainly composed of giants and dwarfs. We 
observed most red giants from their sample, and supplemented this 
set with super-metal rich stars from the Baade's Window region from 
\citet{mcw94} to expand their metallicity range.
Next, we selected red supergiant stars from \citet{whi78} and
\citet{luc89}, to assure good coverage of the stars that dominate
the stellar populations of galaxies with 
ages of 7-12\,Myr -- the so-called red-supergiant-phase. 
In addition, we observed various stars with known metal abundances 
from the catalog of \citet{cay97}. The distribution of our stars 
on the surface gravity log\,$g$ versus effective temperature 
$\rm T_{eff}$ relation is shown in Fig.\,\ref{Fig1}.

The photospheric data for the program stars are summarized in
Table\,\ref{TblStarPars}. About half of them (111 out of 218 stellar
metallicities) were already available from \citet{fab85},
\citet{gor93}, and \citet{wor94b}. The last work offers
re-calculated stellar parameters based on their own spectral line
measurements and calibrations. To ensure that our system is as
close as possible to theirs, we used these new estimates
when available. In addition, we carried out an extensive literature
search for photospheric data for the rest of our program stars. Our
main source was the catalog of [Fe/H] determination of
\citet{cay97} which contains all metallicity estimates in the 
literature up to 1995. Naturally, this is an inhomogeneous 
compilation. In an effort to ``homogenize'' the data as much as 
possible, -- at least in terms of abundance estimate methods -- we 
used spectroscopic determinations when possible. Only did we use 
metallicities based on narrow band photometry when these were the 
only estimates available in the literature. Broad-band photometry 
based metallicities were adopted for six dwarf stars with no other
measurements. 

We made some assumptions when no photospheric parameters could be 
found in the literature: 
(i) solar metallicity was adopted for supergiants in the Galactic 
disk, and 
(ii) [Fe/H]=$-$0.21 was adopted for all bulge stars, following 
\citep{ram00b}. We neglect the 0.3 dex of metallicity dispersion 
in the bulge, until individual estimates become available.
We included in the table some color information -- the reddening 
corrected $(V-K)_0$ -- for stars with unknown spectral class. 

\begin{figure}
\epsscale{0.95}
\plotone{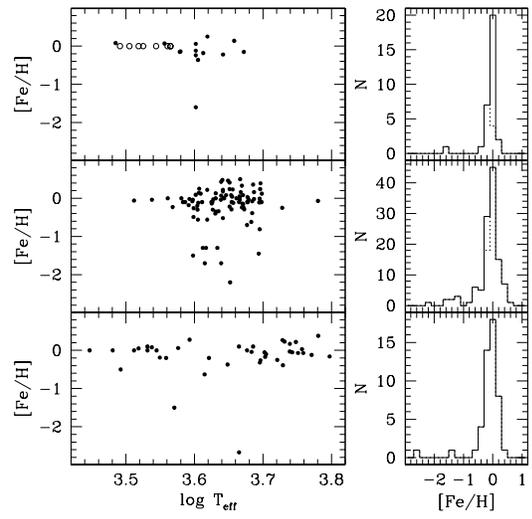}
\caption{Metallicity of library stars [Fe/H] versus the effective 
temperature $\rm T_{eff}$ (left panels), and metallicity histograms
(right) for supergiants, giants, and dwarfs (from top to bottom).
The stars with parameters adopted from the literature are shown with
solid dots, and the stars with asummed parameters are circles. 
Solid-line histograms include all stars with adopted or asummed 
metallicity, and dotted-line histograms omit the asummed values.
\label{Fig4} }
\end{figure}

The metallicities as a function of the luminosity class for stars
in our library are shown in Fig.\,\ref{Fig4}. The giants show the 
largest [Fe/H] spread by far. The supergiants suffer from a pure 
astrophysical constraint: the massive metal poor stars from 
Population II (and possibly III) in our Galaxy exploded as
supernovae a long time ago. The Magellanic Clouds offer a
possibility to complement the library with some supergiants with
1/5 to 1/20 of the solar metallicity \citep{luc98,hil99}. Obtaining 
such spectra is forseen in the next papers of this series.

\section{Index Definitions}

Some spectral features in the near-IR have been observed and
measured, and suitable spectral indices have already been defined
by \citet{kle86}, \citet{ori93}, \citet{doy94}, and \citet{ali95}.
We adopted their definitions for compatibility, with two 
modifications. First, for the NaI and CaI indices of \citet{ali95} 
we used only the two nearest continuum bands. Second, for the CO 
index of \citet{doy94} we carried a polynomial, rather than a 
power-law, fit to the continuum. Experiments show that both these 
changes produce no significant effects, within the errors.

Our spectra do not always span the wavelength range of the original 
photometric CO index of \citet{fro78}, and we were forced to measure 
only narrow spectral indices, including the one defined by
\citet{iva00}. The latter is somewhat intermediate between the 
narrow and broad CO indices, which makes it less sensitive to 
variations of the photospheric transmission. It provides better 
signal-to-noise than the narrower spectroscopic indices.

\begin{figure}
\epsscale{0.95}
\plotone{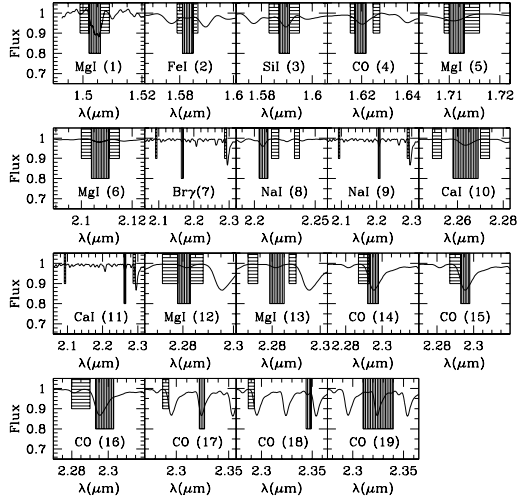}
\caption{Definitions of the spectral indices. The vertically 
shaded area represents the line band, and the horizontally shaded 
areas are the continuum bands. Note that the last index is measured 
after the continuum normalization over the entire K spectra, and 
has no proper continuum band. The bracketed numbers in each panel 
correspond to the index numbering in Table\,\ref{TblIndDef}.
\label{Fig3} }
\end{figure}

Finally, we defined new indices for atomic lines that had not been 
measured before, or where the previous definitions could be affected 
by the loss of the true continuum shape as discussed above, for instance,  
the indices defined by \citet{kle86} where the continuum bands are 
very far apart. All definitions are summarized in 
Table\,\ref{TblIndDef}, and the bandpasses are shown in
Fig.\,\ref{Fig3}. The measured indices for the library stars are 
given in Table\,\ref{TblIndices}. The gaps in the table are due to 
incomplete spectral coverage. 

All indices are in magnitudes:
\begin{equation}
\rm Index = -2.5\times log_{10} (I_{line}/I_{continuum})
\end{equation}
where $\rm I_{line}$ is the flux in the line band, normalized by 
the band width, and $\rm I_{continuum}$ is linear interpolation of 
the continuum flux at the line wavelength. An exception are the CO 
bands for which $\rm I_{continuum}$ is the band width normalized 
flux in the continuum pass band.

\section{Diagnostics of Individual Stars}

\subsection{Stellar Effective Temperature Indicators}

\begin{figure}
\epsscale{0.95}
\plottwo{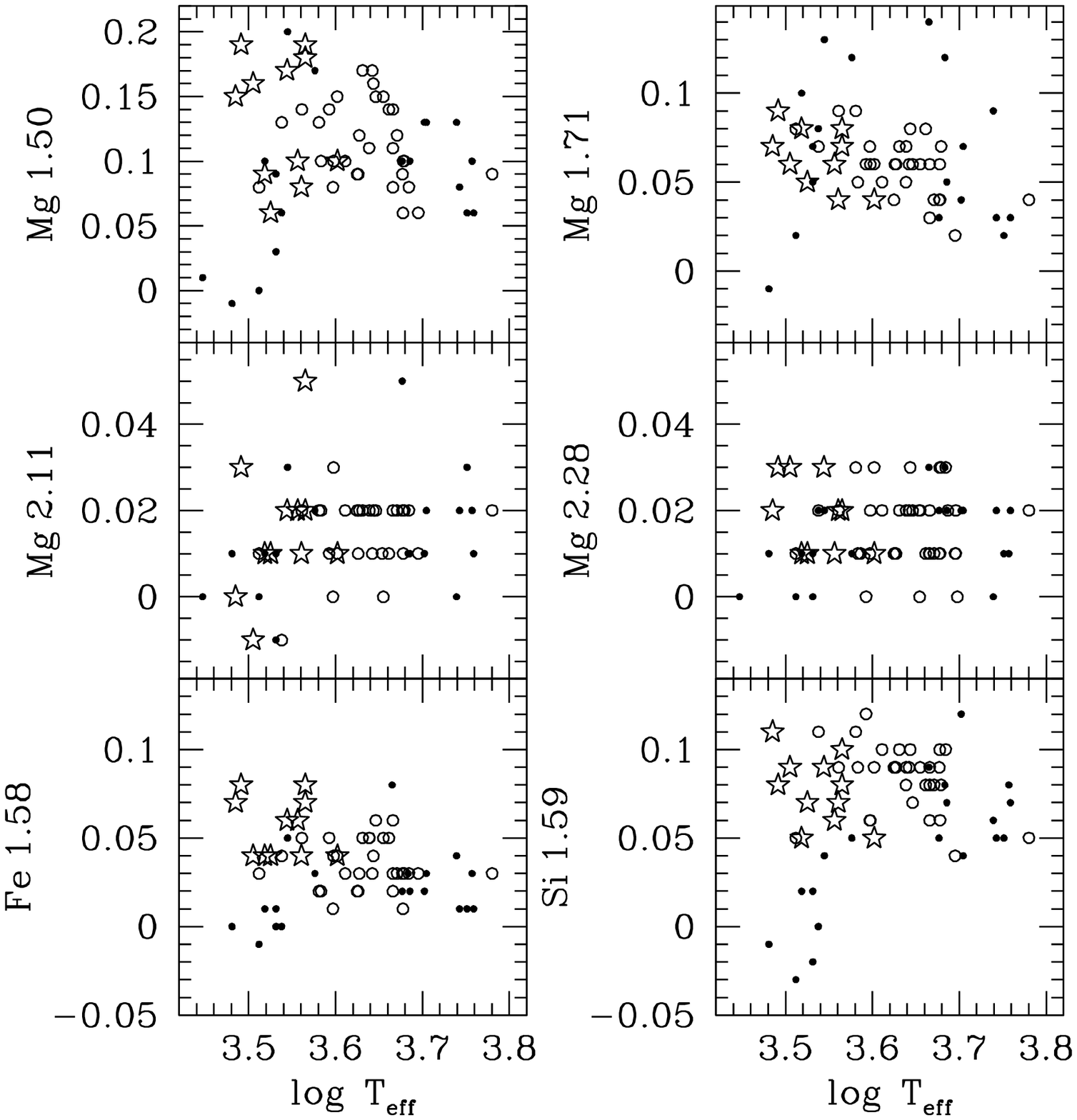}{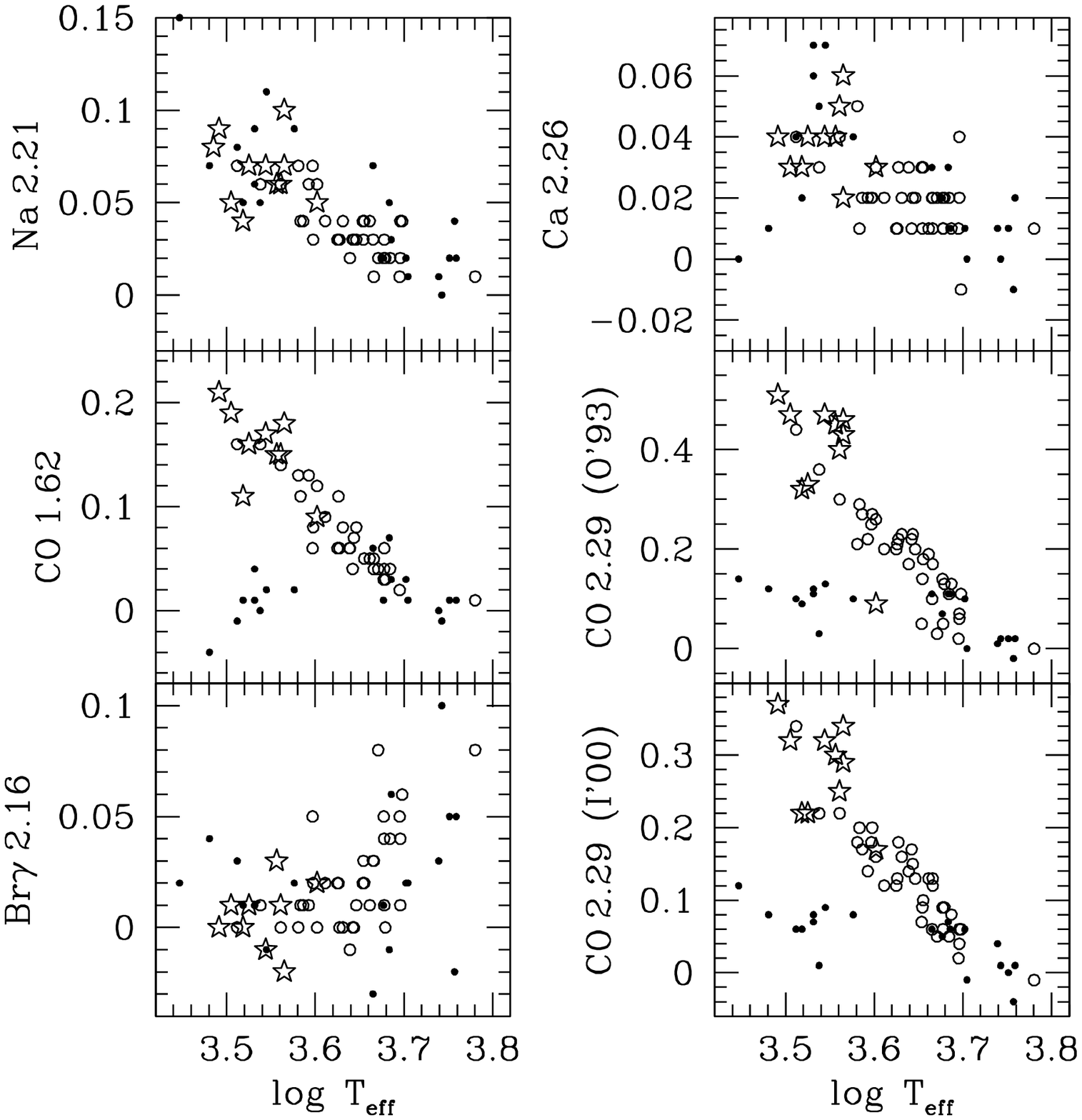}
\caption[Line indices as temperature indicators]{Line indices as
temperature indicators. 
The Si and the CO (upper panel) definitions are from \citet{ori93}, 
the Br$\gamma$ is from \citet{kle86}, the CO (lower panel) is from 
\citet{iva00}, and all MgI and Fe indices are from this work.
Star symbols indicate supergiants, open circles are giants, and solid dots 
represent dwarfs and sub-giants. All indices are in magnitudes. The 
typical measurement error is $\sigma$=0.02 mag. Only stars with 
$-$0.10$\leq$[Fe/H]$\leq$+0.10 are shown.
\label{Fig00} }
\end{figure}

Many near-IR spectral features are good indicators of the stellar 
effective temperature ($\rm T_{eff}$) by themselves, e.g., the CO bands, 
and the NaI, CaI indices \citep{kle86}. 
Fig.~\ref{Fig00} shows the behavior of all 
measured features as function of $\rm T_{eff}$. To minimize 
metallicity effects we only plot stars with 
$-$0.10$\leq$[Fe/H]$\leq$+0.10. The CO bands, Br$\gamma$, Na and Ca 
show stronger temperature dependence than for instance Mg, Fe and 
Si. This behavior has been demonstrated before 
\citep{kle86,ali95,for00}. 

\begin{figure}
\epsscale{0.95}
\plotone{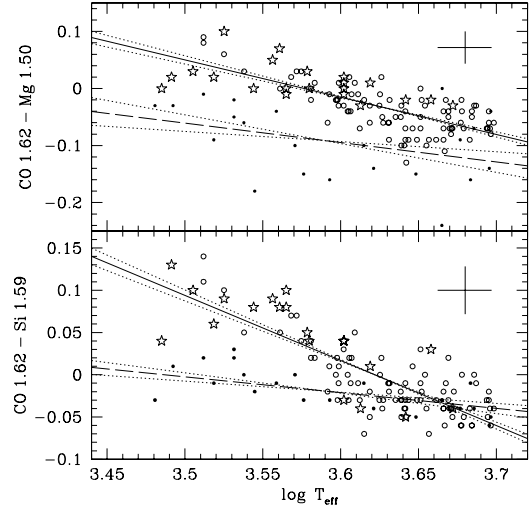}
\caption[Line ratios as temperature indicators]{Line ratios as
temperature indicators. The CO and SiI definitions are from 
\citet{ori93}, and the MgI is from this work. All indices are 
in magnitudes. Star symbols indicate supergiants, open circles 
are giants, and solid dots represent dwarfs and sub-giants. The 
typical $\pm$1$\sigma$ measurement error is shown in the top 
right corner. The best linear fits to supergiants and giant are 
shown with a solid line, and the best fit to dwarfs and subgiants 
is a dashed line. The dotted lines represent $\pm$1$\sigma$ errors 
in the slopes.
\label{Fig26a} }
\end{figure}

Some line ratios, for instance EW(CO\,1.62)/EW(SiI\,1.59) 
\citep{ori93,dal96,for00},
are better temperature indicators than individual 
lines because the division cancels out any additional luminosity, 
metallicity, or reddening effects. Based on the indices of 109 stars from 
our sample (Fig.\,\ref{Fig26a}) we find the following relation for 
giants and supergiants with $\rm T_{eff}$$\leq$5000 K:
\begin{equation}
\rm (CO\,1.62 - SiI\,1.59)=(2.79\pm0.19)-(0.77\pm0.05)\times log\,T_{eff}
\end{equation}
Here we used the index definitions of \citet{ori93}. 
The root-mean-square (hereafter, r.m.s.) of 
the relation is 0.03 mag, which corresponds to $\sim$300 K for the 
inverse equation, close to the typical observational errors of 0.02 
mag. From indices of 23 dwarfs and sub-giants, again with 
$\rm T_{eff}$$\leq$5000 K: 
\begin{equation}
\rm (CO\,1.62 - SiI\,1.59)=(0.65\pm0.20)-(0.19\pm0.05)\times log\,T_{eff}
\end{equation}
where the r.m.s is 0.02 mag, corresponding to $\sim$800 K. The 
slopes of the relations are significantly different, and the 
loci of the two groups overlap only for hot stars where both 
spectral features are weak and the relative errors increase.

We derived similar relations using the $1.50\,\rm\mu m$ MgI feature.
For 107 giants and supergiants:
\begin{equation}
\rm (CO\,1.62 - MgI\,1.50)=(2.34\pm0.21)-(0.65\pm0.06)\times log\,T_{eff}
\end{equation}
The CO index is defined by \citet{ori93} and the MgI index is 
defined in this work. The r.m.s. is 0.03 mag or $\sim$350 K. 
For 23 dwarfs and sub-giants: 
\begin{equation}
\rm (CO\,1.62 - MgI\,1.50)=(1.13\pm0.59)-(0.34\pm0.16)\times log\,T_{eff}
\end{equation}
with r.m.s. of 0.06 mag or $\sim$1500 K. 
The relations for dwarfs and sub-giants are worse than for giants 
and supergiants because of both weaker spectral lines and 
systematically fainter targets. Note that even though we have not 
restricted the abundances, the majority of stars used to derive 
the equations above have near-solar metallicities 
($-$0.2$\leq$[Fe/H]$\leq$+0.2) and abundance ratios.

\subsection{Two-dimensional Spectral Classification -- Luminosity 
Class Indicators}

\begin{figure}
\epsscale{0.95}
\plotone{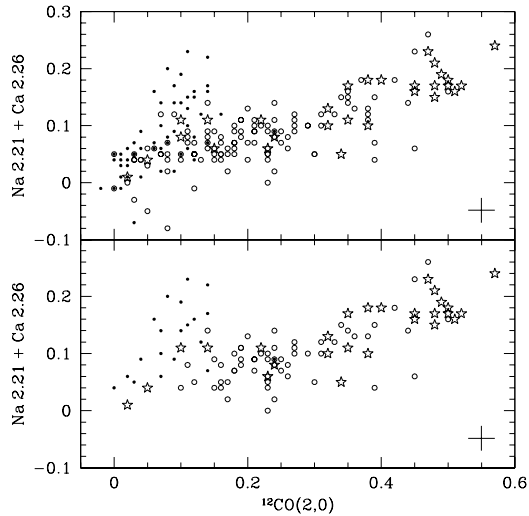}
\caption[Two-dimensional Spectral Classification]{Two-dimensional 
Spectral Classification.
Comparison of the strengths of Na and Ca features, with the 
$2.29\rm\mu m$ CO bandhead absorption, following \citet{kle86}. 
All index definitions are from there and the indices are in magnitudes.
Star symbols indicate supergiants, open circles are giants, and solid 
dots represent dwarfs and sub-giants. The typical $\pm$1$\sigma$ 
observational uncertainties are shown in the bottom right corner.
The top panel includes all stars, the bottom panel shows only 
stars with [Fe/H]$\geq$$-$0.5 and $\rm T_{eff}\leq4500$ K.
\label{Fig24a} }
\end{figure}

The two-dimensional spectral classification requires an indicator
for the intrinsic luminosity of the stars. \citet{kle86} 
demonstrated that Na, Ca and Br$\gamma$ can be used to discriminate 
between stars of different luminosity classes (see their Fig. 7).
We verified their result, excluding the Br$\gamma$ index 
(Fig.~\ref{Fig24a}) to minimize the uncertainties related to the 
removal of the intrinsic Br$\gamma$ absorption in the telluric 
standards (Sec.\,3). The (super)giant versus dwarf separation is 
relatively small, compared with the typical observational 
uncertainties. The situation improves if we constrain the sample 
only to metal rich stars with [Fe/H]$\geq$$-$0.5 (bottom panel), 
minimizing the relative errors. Clearly, this diagnostic relationship 
imposes a high demand on the data quality.

\begin{figure}
\epsscale{0.95}
\plottwo{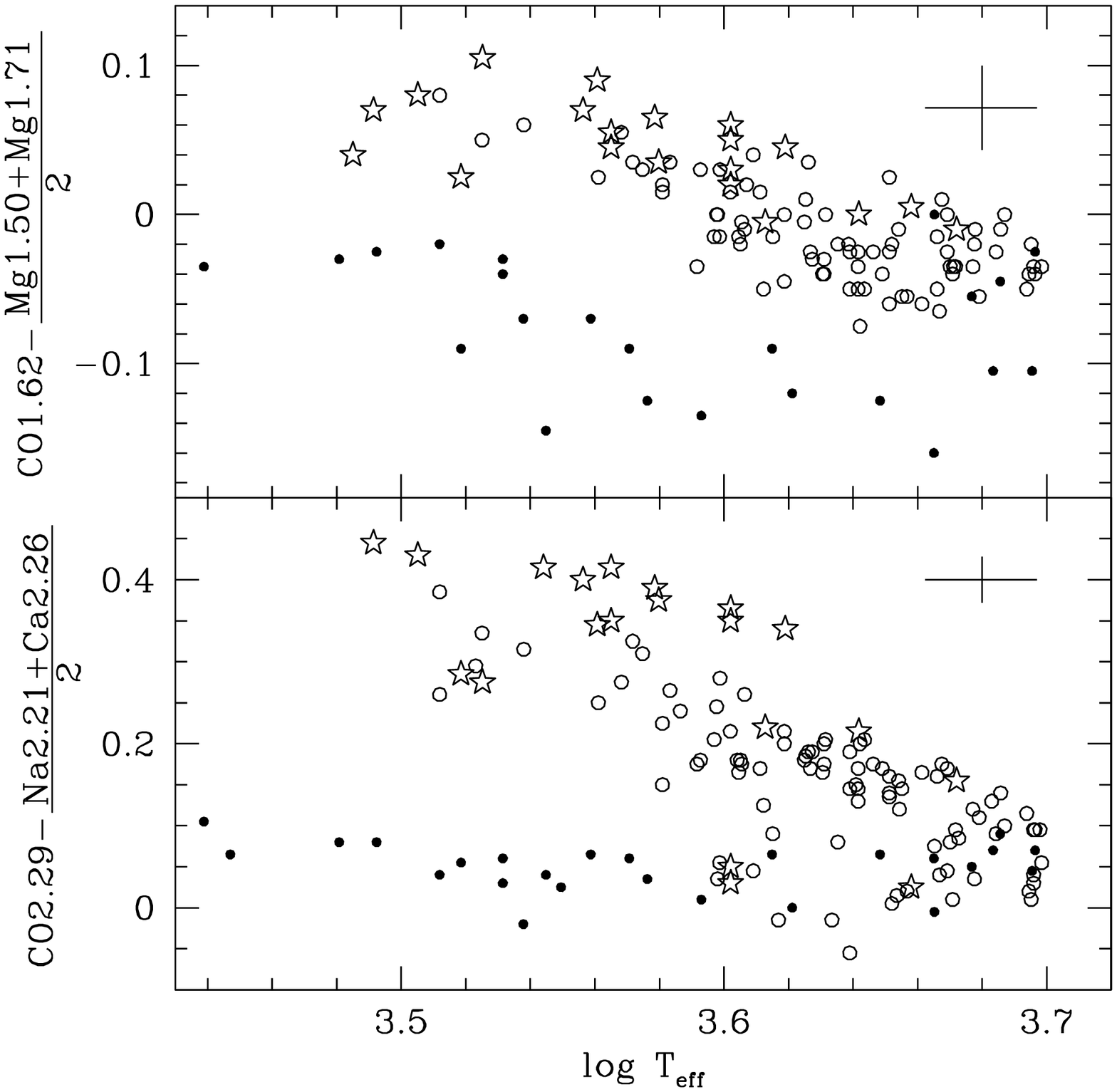}{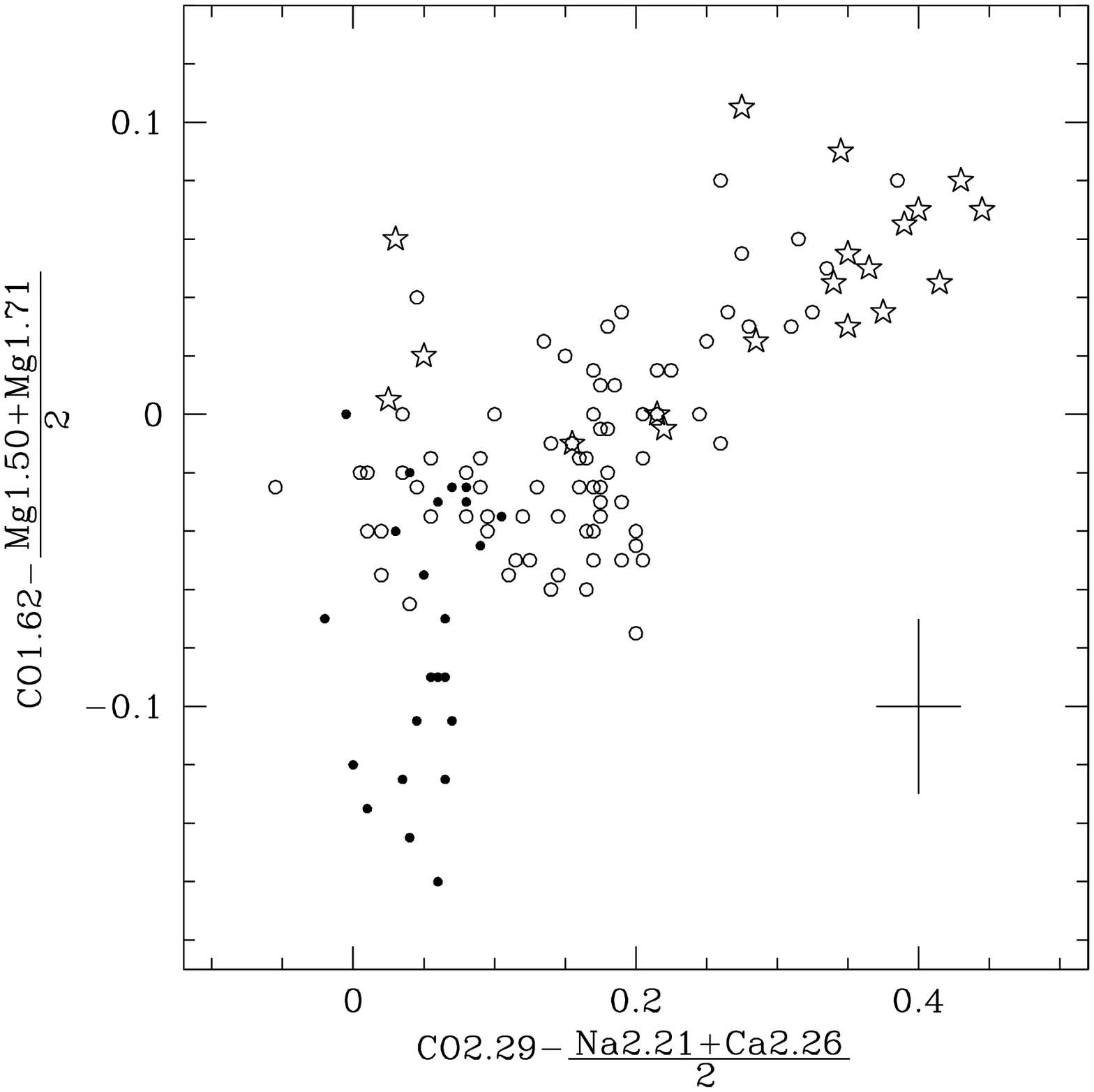}
\caption[Two-dimensional Spectral Classification]{Two-dimensional 
Spectral Classification with H-band features (top) and K-band 
features (bottom). All indices are in magnitudes.
The stellar effective temperature $\rm T_{eff}$ is used on the 
left plot, and only pure observables are used on the right.
Star symbols indicate supergiants, open circles are giants, and 
solid dots represent dwarfs and sub-giants. The typical 
$\pm$1$\sigma$ measurement error is shown in the bottom right 
corner.
The CO bands are defined by \citet{ori93}, the Ca and Na indices 
are from \citet{ali95}, and the Mg definitions are from this work.
\label{Fig24b} }
\end{figure}

\citet{ram97} proposed to use the 
log\{EW(CO\,2.29)/[EW(NaI\,2.21)+EW(CaI\,2.26)]\} ratio plotted 
against $\rm T_{eff}$ to separate 
giants from dwarfs (see their Fig. 11). Our data confirm this 
result (Fig.\,\ref{Fig24b}, bottom left), even for higher 
$\rm T_{eff}$ than \citet{ram97} because of the improved S/N and
spectral resolution. We obtain similar separation with Mg features 
(Fig.\,\ref{Fig24b}, top left). This methods use a physical 
quantity -- $\rm T_{eff}$ -- that requires some calibration to be 
derived from observables such as $(V-K)_0$ or some spectral 
feature. To avoid this additional step, we combined the two 
ratios -- log\{EW(CO\,2.29)/[EW(NaI\,2.21)+EW(CaI\,2.26)]\} and 
log\{EW(CO\,1.62)/[EW(MgI\,1.50)+EW(MgI\,1.71)]\}, -- and found 
that we can still achieve separation for most of the stars. 

The signal-to-noise ratio needed to use the diagnostics discussed
here depends on the temperature of the stars. It is safe to assume 
that S/N$\sim$30 is necessary for K5-M stars, and it increases
up to 50 for early K stars. The lines become too weak to implement
these techniques for stars with effective temperatures hotter than
$4500-4800\,$K. An additional limitation comes from the metal 
abundance -- it is more difficult to separate stars of different 
luminosity class with low metallicity than with high metallicity, 
because in the former the lined are weaker, and the relative 
uncertainties are higher. However, our library does not offer
sufficient abundance range to quantify this effect.

\section{Summary}

We have assembled a library of moderately high-resolution
($\approx$\,2000-3000) $H$ (1.6\,$\mu$m) and $K$ (2.2\,$\mu$m)
spectra of 218 red stars, mostly supergiants and giants. The 
majority of these stars have well-known photospheric parameters 
from high-resolution optical spectroscopy. 
These stars dominate the near-IR emission in both starburst and 
elliptical galaxies. Our library covers a range of effective 
temperatures, and metal abundances from [Fe/H]$\sim$$-$2.2 to 
+0.3. This library will offer a unique opportunity to study 
directly the most obscured stellar populations in starburst 
galaxies, as well as in the center of the Milky Way.

Although the main motivation behind the creation of this library 
is the study of unresolved extragalactic stellar populations, 
the obtained spectra can be used to derive parameters of 
individual stars. We calibrated some line ratios as indicators 
of the stellar effective temperature. Finally, we demonstrated how
some diagnostic relationships can distinguish (super)giants from 
dwarf stars.

\acknowledgments

This research has made use of the SIMBAD database, operated 
at CDS, Strasbourg, France. The authors were supported by NSF 
grant AST 95-29190.
We are grateful to the anonymous referee for the suggestions 
that helped to improve the paper.

\begin{deluxetable}{l@{}c@{ }c@{}c@{}c}
\tablewidth{0pc}
\tablecaption{Near-IR spectral libraries. This work is
included for comparison.
\label{TblLibRev}}
\footnotesize
\tablehead{
\multicolumn{1}{c}{Spectral Library}& 
\multicolumn{1}{c}{$\lambda$}&
\multicolumn{1}{c}{Sp. Type and}& 
\multicolumn{1}{c}{No.}& 
\multicolumn{1}{c}{Spectral}\\
\multicolumn{1}{c}{Reference}& 
\multicolumn{1}{c}{$\micron$}&
\multicolumn{1}{c}{Lum. Class range}& 
\multicolumn{1}{c}{Stars}& 
\multicolumn{1}{c}{Resolution}}
\startdata
\citet{joh70} & 1.2-2.5   & A-M, I-V &  32 & 550 \\
\citet{kle86} & 2.0-2.5   & F-M, I-V &  26 & 2500-3100\\
\citet{lan92} & 1.4-2.5   & O-M, I-V &  56 & 550 \\
\citet{ori93} & 1.5-1.7   & G-M, I-V &  40 & 1500\\
\citet{ali95} & 2.0-2.4   & F-M, V   &  33 & 1380\\
\citet{wal96} & 2.02-2.41 & G-M, I-V &  12 & 45000    \\
\citet{dal96} & 1.57-1.64 & O-M, I-V &  37 & 1500-2000\\
\citet{han96} & 2.0-2.4   & O-B, I-V & 180 & 800-3000 \\
\citet{ram97} & 2.19-2.34 & K-M, III &  43 & 1380,4830\\
\citet{wal97} & 2.0-2.4   & O-M, I-V & 115 & 3000\\
\citet{mey98} & 1.5-1.7   & O-M, I-V &  85 & 3000\\
\citet{joy98} & 1.0-2.5   & K-M, III & 29 & 500-1500\\
\citet{for00} & 1.90-2.45 & G-M, I-III & 31 & 830,3000\\
\citet{wal00} & 1.05-1.34 & O-M, I-V &  88 & 3000\\
\citet{lan00} & 0.5-2.5   & K-M, I-III & 77 & 1100\\
\citet{hic00} & 1.08-1.35 & O-M, I-V & 105 & 650\\
This work     & 1.48-2.45 & G-M, I-V & 218 & 2000-3000\\
\enddata
\end{deluxetable}{}

\begin{deluxetable}{c@{ }c@{ }c@{ }c@{ }c@{ }c@{ }c}
\tablewidth{0pc}
\tablecaption{Log of spectroscopic observations.
\label{TblObsLog}}
\footnotesize
\tablehead{
\multicolumn{1}{c}{Date} & 
\multicolumn{1}{c}{Site} &
\multicolumn{1}{c}{Star} &
\multicolumn{1}{c}{Std.} & 
\multicolumn{1}{c}{Sp.T.} &
\multicolumn{1}{c}{Gr.} &  
\multicolumn{1}{c}{Range}\\
\multicolumn{1}{c}{(1)} &  
\multicolumn{1}{c}{(2)} &  
\multicolumn{1}{c}{(3)} &
\multicolumn{1}{c}{(4)} &  
\multicolumn{1}{c}{(5)} &  
\multicolumn{1}{c}{(6)} &
\multicolumn{1}{c}{(7)} }
\startdata
% This table was made by mklog03.c from all_stars06.txt (or a later version).
951209&61&  S074439&  S074506  &A3 V&600&HK\\
951209&61&  S060890&   HR3348  &A0 V&600&HK\\
951209&61&  S027179&   HR3592  &A2 V&600&HK\\
951209&61&    VYLeo&   HR4244  &A3 V&600&H\\
951210&61&  S074175&   HR0277  &A2 V&600&HK\\
951210&61&  S109474&   HR0081  &A0 V&600&HK\\
951210&61&  S109471&   HR0081  &A0 V&600&HK\\
951210&61&  S074883&   HR0669  &A1 V&600&2.30-2.40\\
\enddata
\end{deluxetable}

\begin{deluxetable}{r@{ }r@{ }r@{ }r@{ }c@{ }c@{ }c@{ }c@{ }c@{ }c}
\tablecaption{Adopted photospheric parameters of the stars. The
complete table is presented in the electronic edition only.
\label{TblStarPars}}
\footnotesize
\tablehead{
\multicolumn{5}{c}{Stellar ID\tablenotemark{a}} &
\multicolumn{1}{c}{Sp.} &
\multicolumn{4}{c}{Photospheric Parameters\tablenotemark{b}} \\
\multicolumn{1}{c}{HR} &
\multicolumn{1}{c}{HD} &
\multicolumn{1}{c}{SAO} &
\multicolumn{1}{c}{BD} &
\multicolumn{1}{c}{Other} &
\multicolumn{1}{c}{Class} &
\multicolumn{1}{c}{$\rm T_{eff}$} &
\multicolumn{1}{c}{log\,$g$ } &
\multicolumn{1}{c}{[Fe/H]} &
\multicolumn{1}{c}{Method\tablenotemark{c}}\\
\multicolumn{1}{c}{(1)} &  
\multicolumn{1}{c}{(2)} &  
\multicolumn{1}{c}{(3)} &
\multicolumn{1}{c}{(4)} &  
\multicolumn{1}{c}{(5)} &  
\multicolumn{1}{c}{(6)} &
\multicolumn{1}{c}{(7)} &
\multicolumn{1}{c}{(8)} &  
\multicolumn{1}{c}{(9)} &  
\multicolumn{1}{c}{(10)} }
\startdata
\multicolumn{10}{c}{}\\
\multicolumn{10}{c}{SUPERGIANTS}\\
    &104452& 82106& 22 2430&       1Com&  G0     II(32)&   -1    &99.99    & 0.00(1) &-\\
7063&173764&142618& -4 4582& $\beta$Sct&  G4    IIa(32)& 4700(2) & 0.94(2) &-0.15(2) &s\\
7314&180809& 68065& 37 3398&$\theta$Lyr&  K0     II(32)& 4550(2) & 1.77(2) & 0.14(2) &s\\
6713&164349&103285& 16 3335&      93Her&K0.5     II(32)& 4383(2) & 1.80(2) &-0.22(2) &s\\
    &193515& 69825& 37 3882&   PPM84702&  K1     II(33)&   -1    &99.99    & 0.00(1) &-\\
8465&210745& 34137& 57 2475& $\zeta$Cep&K1.5     Ib(32)& 4159(2) & 0.88(2) & 0.25(2) &s\\
\enddata
\end{deluxetable}

\begin{deluxetable}{c@{ }c@{ }c@{ }c@{ }c@{ }c@{ }c@{ }c@{}c}
\tablewidth{0pc}
\tablecaption{Definitions of spectral bands.
\label{TblIndDef}}
\footnotesize
\tablehead{
\multicolumn{1}{c}{No.}&
\multicolumn{1}{c}{Species}&
\multicolumn{2}{c}{Line\tablenotemark{a}}&
\multicolumn{2}{c}{Cont. 1\tablenotemark{a}}&
\multicolumn{2}{c}{Cont. 2\tablenotemark{a}}&
\multicolumn{1}{c}{Ref.}\\
\multicolumn{1}{c}{}&
\multicolumn{1}{c}{}&
\multicolumn{1}{c}{$\lambda_c,\AA$}&
\multicolumn{1}{c}{$\Delta\lambda,\AA$}&
\multicolumn{1}{c}{$\lambda_c,\AA$}&
\multicolumn{1}{c}{$\Delta\lambda,\AA$}&
\multicolumn{1}{c}{$\lambda_c,\AA$}&
\multicolumn{1}{c}{$\Delta\lambda,\AA$}&
\multicolumn{1}{c}{}}
\startdata
 1 & MgI  (1.50$\micron$) & 15040   &40 & 15005   &30 & 15075   &30 & 1\\
 2 & FeI  (1.58$\micron$) & 15830   &40 & 15800   &20 & 15857.5 &15 & 1\\
 3 & SiI  (1.59$\micron$) & 15890   &40 & 15850   &40 & 15930   &40 & 3\\
 4 & CO   (1.62$\micron$) & 16197.5 &45 & 16160   &30 & 16270   &30 & 3\\
 5 & MgI  (1.71$\micron$) & 17115   &30 & 17092.5 &15 & 17145   &30 & 1\\
 6 & MgI  (2.11$\micron$) & 21075   &70 & 21020   &40 & 21130   &40 & 1\\
 7 & Br$\gamma$ (2.16$\micron$) & 21662.5 &47 & 20929  &44 & 22899  &52 & 2\\
 8 & NaI  (2.21$\micron$) & 22075   &70 & 22170   &60 & 22350   &40 & 4\\
\enddata
\end{deluxetable}

\begin{deluxetable}{l@{}c@{}c@{}c@{}c@{}c@{}c@{}c@{}c@{}c@{}c@{}c@{}c@{}c@{}c@{}c@{}c@{}c@{}c@{}c}
\tablewidth{0pc}
\tablecaption{Spectral indices of features of interest.
\label{TblIndices}}
\footnotesize
\tablehead{
\multicolumn{1}{c}{Star ID}&
\multicolumn{1}{c}{Mg}&
\multicolumn{1}{c}{Fe}&
\multicolumn{1}{c}{Si}&
\multicolumn{1}{c}{CO}&
\multicolumn{1}{c}{Mg}&
\multicolumn{1}{c}{Mg}&
\multicolumn{1}{c}{$\rm Br\gamma_k$}&
\multicolumn{1}{c}{Na}&
\multicolumn{1}{c}{Na}&
\multicolumn{1}{c}{Ca}&
\multicolumn{1}{c}{Ca}&
\multicolumn{1}{c}{Mg}&
\multicolumn{1}{c}{Mg}&
\multicolumn{1}{c}{CO}&
\multicolumn{1}{c}{CO}&
\multicolumn{1}{c}{CO}&
\multicolumn{1}{c}{CO}&
\multicolumn{1}{c}{CO}&
\multicolumn{1}{c}{CO}}
\startdata
\multicolumn{20}{c}{}\\
\multicolumn{20}{c}{SUPERGIANTS}\\
  SAO082106& 0.04& 0.01& 0.04& 0.01& 0.00& 0.00& 0.08& 0.01& 0.01& 0.01& 0.00& 0.01& 0.02& 0.02& 0.02& 0.01& 0.01& 0.01& 0.01\\
     HR7063& 0.06& 0.04& 0.07& 0.03& 0.02& 0.01& 0.07& 0.00& 0.03& 0.01& 0.03& 0.01& 0.01& 0.16& 0.15& 0.09& 0.14& 0.01& 0.05\\
     HR7314& 0.08& 0.03& 0.03& 0.06& 0.03& 0.02& 0.01& 0.03& 0.03& 0.04& 0.05& 0.00& 0.01& 0.06& 0.10& 0.11& 0.20& 0.02& 0.07\\
     HR6713& 0.07& 0.03& 0.10& 0.05& 0.03& 0.02& 0.04& 0.00& 0.04& 0.01& 0.02& 0.02& 0.02& 0.22& 0.23& 0.14& 0.19& 0.01& 0.06\\
  SAO069825& 0.07& 0.02& 0.02& 0.02& 0.03& 0.02& 0.07& 0.02& 0.02& 0.02& 0.02& 0.01& 0.01& 0.03& 0.05& 0.06& 0.12& 0.00& 0.04\\
     HR8465& 0.10& 0.04& 0.10& 0.11& 0.03& 0.01& 0.03& 0.03& 0.06& 0.01& 0.04& 0.02& 0.01& 0.36& 0.38& 0.25& 0.31& 0.07& 0.09\\
\enddata
\end{deluxetable}

\newpage

\end{document}